\pdfoutput=1 
\documentclass[submission, Proceedings]{SciPost}

\hypersetup{
    colorlinks,
    linkcolor={red!50!black},
    citecolor={blue!50!black},
    urlcolor={blue!80!black}
}

\usepackage[bitstream-charter]{mathdesign}
\DeclareSymbolFont{usualmathcal}{OMS}{cmsy}{m}{n}
\DeclareSymbolFontAlphabet{\mathcal}{usualmathcal}

\begin{document}

\begin{center}{\Large \textbf{
Particle Physics with the Pierre Auger Observatory\\
}}\end{center}

\begin{center}
Matias Perlin\textsuperscript{1,2} on behalf of the Pierre Auger Collaboration\textsuperscript{3$\star$}
\end{center}

\begin{center}
{\bf 1} Instituto de Tecnologías en Detección y Astropartículas (CNEA, CONICET, UNSAM), Buenos Aires, Argentina \\
{\bf 2} Karlsruhe Institute of Technology (KIT), Institute for Astroparticle Physics, Karlsruhe, Germany \\
{\bf 3} Observatorio Pierre Auger, Av. San Martín Norte 304, 5613 Malargüe, Argentina \\
Full Author list: \url{https://www.auger.org/archive/authors_2021_04.html}
\\
%
$\star$ \url{spokespersons@auger.org}
\end{center}

\begin{center}
\today
\end{center}


\definecolor{palegray}{gray}{0.95}
\begin{center}
\colorbox{palegray}{
  \begin{tabular}{rr}
  \begin{minipage}{0.1\textwidth}
    \includegraphics[width=22mm]{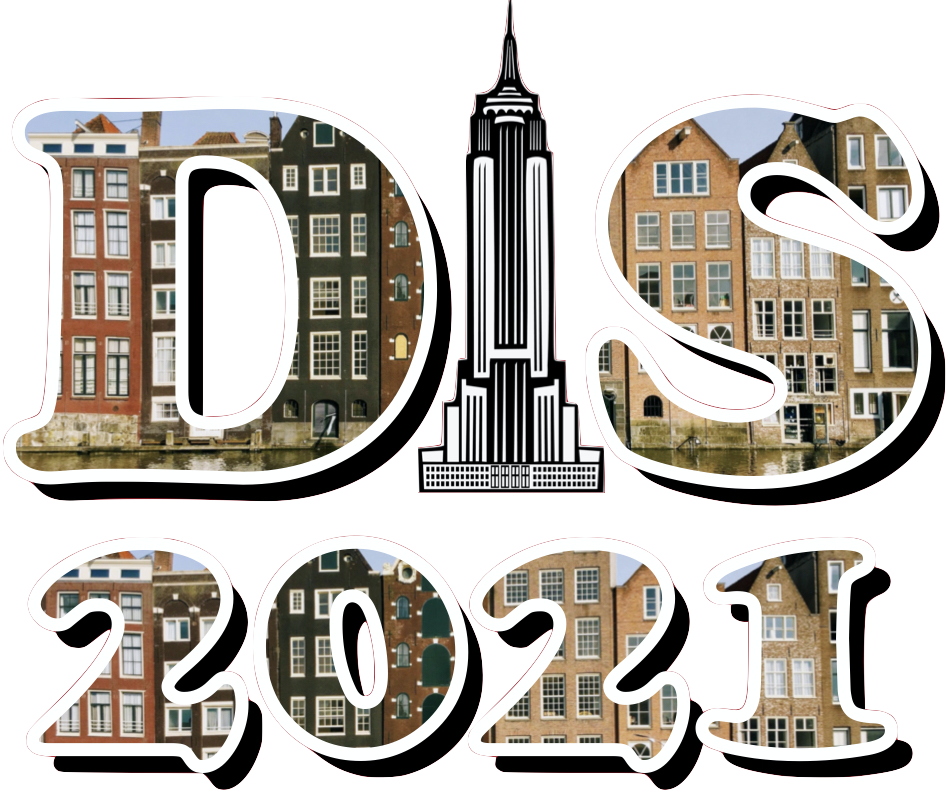}
  \end{minipage}
  &
  \begin{minipage}{0.75\textwidth}
    \begin{center}
    {\it Proceedings for the XXVIII International Workshop\\ on Deep-Inelastic Scattering and
Related Subjects,}\\
    {\it Stony Brook University, New York, USA, 12-16 April 2021} \\
    \doi{10.21468/SciPostPhysProc.?}\\
    \end{center}
  \end{minipage}
\end{tabular}
}
\end{center}

\section*{Abstract}
{\bf
%
The Pierre Auger Observatory is the largest extensive air shower
detector. Based on a hybrid system, this experiment measures the longitudinal shower development 
and the particles at the ground. 
This detection system allows the extraction of the p-air cross-section at energies much
higher than the ones accessible by current colliders. 
It is also possible to test hadronic interaction models using correlations between different air shower
observables, like the depth of shower maximum and the muon number at the ground and their fluctuations.
Thanks to the low energy extension of the Pierre Auger Observatory, the muon deficit in air shower
simulations can be addressed over almost three decades at the highest energies.
}

\section{Introduction}
\label{sec:intro}
The search of new particle interaction properties is closely related to the energy of the interactions.
Human-made accelerators are being continuously improved reaching p-p interactions at a center of mass 
energy $\sqrt s=13$ TeV  at the LHC.
However, interactions up to $\sqrt s \sim 200$ TeV   happen in the upper atmosphere when 
ultra-high energy cosmic rays collide with air nuclei. 
The direct detection of these cosmic rays is impossible to achieve with good statistics since its flux is extremely low. 
Fortunately, these collisions produce secondary particles that interact again creating a huge
cascade of particles known as extensive air showers (EAS) of cosmic rays. 
The fluorescence light from the de-excitation of the atmospheric nitrogen  
and the particles at the ground can be measured to infer the properties of the initial interaction.
In this way, besides addressing astrophysics questions, EAS measurements provide access to hadronic 
interactions at energies unreached by current experiments.

\section{The Pierre Auger Observatory}
\label{sec:pao}
The Pierre Auger Observatory \cite{PierreAuger:2015eyc}, located near the city of Malargüe in Argentina, 
is the largest facility for ultra-high energy cosmic-ray detection in the world.
It provides a hybrid detection of EAS combining two independent techniques.
The longitudinal development of air showers in
the atmosphere is observed by the Fluorescence Detector (FD), and the lateral
distribution of particles at the ground is measured by the Surface
Detector (SD). 

The FD is comprised of 27 telescopes spread over four sites which measure the fluorescence light produced 
in the atmosphere by the charged particles of the EASs through the 
excitation of nitrogen molecules.
This light is proportional to the energy deposited by the shower particles, 
and the integral of this profile allows a quasi-calorimetric measurement of the cosmic ray energy 
when correcting for the missing energy carried by mostly muons and neutrinos.
The duty cycle is limited to 15\% since the FD requires clear and moonless nights to operate. 

The SD consists of 1660 independent water-Cherenkov detectors 
(WCD) with a duty cycle close to 100\%  equipped with photomultipliers to detect the Cherenkov light emitted in
water by electromagnetic particles and muons. 
The standard array (SD-1500) is distributed on a triangular grid of 1500 m spacing and covers about 3000 km$^2$. 
Moreover, a denser deployment of stations named Infill array (SD-750)  with 750 m separation within the main 
array allows for a lower energy threshold.
The SD cannot measure the primary energy directly but using a cross-calibration with the FD, 
it is possible  to estimate the energy for SD-only events.

A major upgrade called AugerPrime \cite{PierreAuger:2016qzd} is taking place in the Pierre Auger Observatory.
For the purpose of improving the mass composition estimation, each SD station is being  equipped with
 a scintillator layer (SSD) on the top which has a different response to the muonic and electromagnetic components.
Moreover, the Auger Muons and Infill for the Ground Array (AMIGA) upgrade allows  direct detection of the muon content  
thanks to the so-called Underground Muon Detectors (UMD). 
These muon detectors are being deployed next to each surface detector of the Infill Array (SD-750) and 
buried at 2.3 m underground, giving a muon energy threshold of $1~\mathrm{GeV}$.

\section{Mass composition}
\label{sec:Spectrum_MassComp}
\begin{figure}[h!]
\centering
\includegraphics[width=0.99\textwidth]{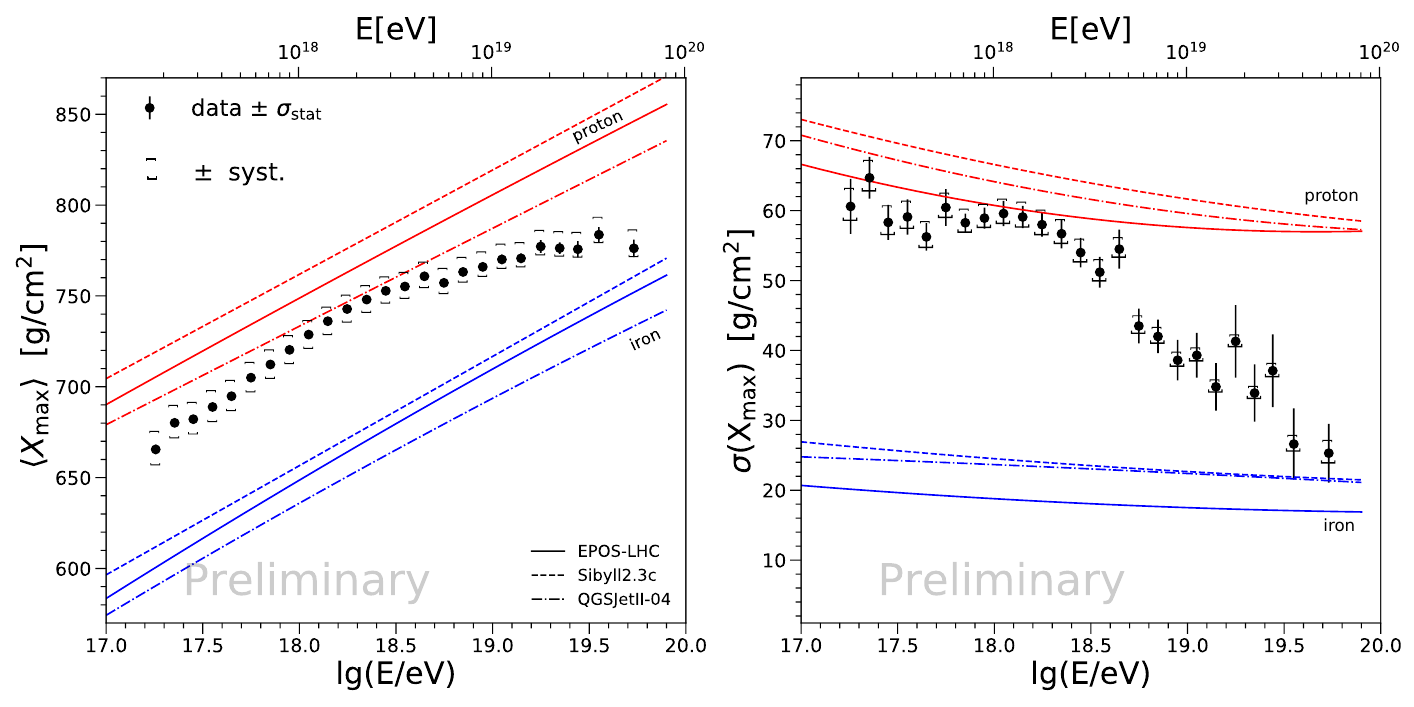}
\vspace{-2mm}
\caption{The energy evolution of the mean of $X_{\rm{max}}$ (left) and $\sigma(X_{\rm{max}})$ (right)
measured by the Auger Observatory compared to proton and iron shower simulations \cite{Yushkov:2020nhr}.}
\label{Fig:mass}
\end{figure}

One of the main goals  of the Pierre Auger Observatory is to understand the mass composition of 
cosmic rays due to the link with its sources.
An important observable is the depth of the
shower maximum $X_{\rm{max}}$, i.e., the atmospheric depth of the maximum energy deposit by shower particles, 
measured as a atmospheric column density in units of g/cm$^2$.
The $X_{\rm{max}}$ is sensitive to mass composition since showers from heavier
nuclei develop at higher altitudes in the atmosphere (lower $X_{\rm{max}}$) and 
their profiles fluctuate less according to the nuclear superposition model, 
while showers from lighter nuclei develop deeper in the atmosphere (larger $X_{\rm{max}}$) with larger fluctuations.
In hybrid events, the $X_{\rm{max}}$ is measured directly by the FD and 
the latest results are shown in Fig. \ref{Fig:mass}, where the $X_{\rm{max}}$ distribution  
and its fluctuations $\sigma(X_{\rm{max}})$ are presented as a function of energy.
All models show that the mass composition is lighter at low energies  and after the ankle 
of the spectrum ($ \sim 6 \times 10^{18}~\mathrm{eV}$), it becomes heavier.
However, the composition at each energy cannot be established since each hadronic model gives a different prediction. 

\section{The p-air cross section}
\label{sec:p-Air}
The analysis of the  $X_{\rm{max}}$ distribution  allows one to make direct measurements 
and tests of hadronic properties such as the p-air cross section $\sigma_{\rm{p-air}}$.
The  $X_{\rm{max}}$ can be expressed as $X_{\rm{max}} = X_1+\Delta$, 
where $X_1$ is the depth of the first interaction and $\Delta$ the air shower development length itself.
As the lightest elements dominate the upper tail of the  $X_{\rm{max}}$ distribution and in turn, 
the $X_1$ distribution is an exponential,
the $\text{d}N/\text{d}X_{\rm{max}}$ tail goes as exp$(-X_{\rm{max}}/\Lambda_{\eta})$ 
with $\Lambda_{\eta}^{-1} \simeq \sigma_{\rm{p-air}}$.
The Pierre Auger Collaboration has measured the $\sigma_{\rm{p-air}}$ in the two energy intervals 
$\log_{10}(E/\text{eV})= \allowbreak [17.8-18] \text{ and } \allowbreak [18-18.5]$, 
where the showers come  mostly from protons. 
Using the Glauber theory, it is also possible to obtain the inelastic p-p cross section $\sigma_{\rm{p-p}}^{\rm{inel}}$ 
from $\sigma_{\rm{p-air}}$  at center of mass energies $\sqrt s=38.7~\mathrm{TeV}$ 
and $55.5~\mathrm{TeV}$. 
These results agree with extrapolations from LHC energies.
Details of these analyses can be found in \cite{PierreAuger:2012egl,Ulrich:2015yoo}.

\section{Number of muons}
\label{sec:Nmu}
Over recent decades, different experiments have reported that the number of muons present in air showers 
is significantly larger than  the one expected from simulation.
The lack of muons given by hadronic interaction models is known as \emph{muon deficit}.
The Pierre Auger Observatory has different observables sensitive to the muonic content of air showers.

\begin{figure}[ht!]
\centering
\includegraphics[width=0.99\textwidth]{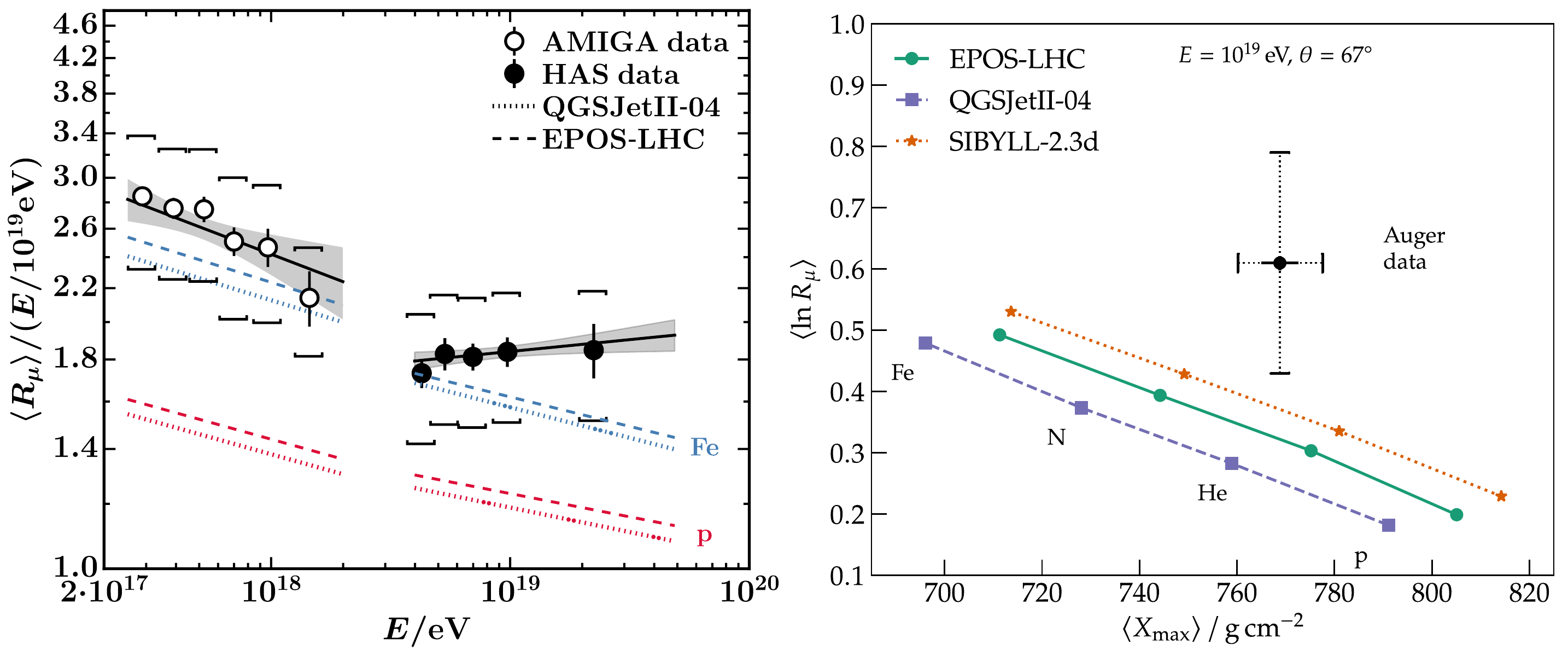}
\vspace{-2mm}
\caption{ 
Left: The energy evolution of the average muon content $R_{\mu}$ measured by AMIGA (white circles) 
and by the SD using horizontal air showers (black circles) \cite{Cazon:2019ywa}.
Right: Average logarithmic muon content as a function of the average shower depth  $X_{\rm{max}}$ \cite{PierreAuger:2021qsd}.
The values expected from air-showers simulations are also shown.
}
\label{Fig:Rmu2}
\end{figure}

One method is to measure the muon content directly using highly inclined showers 
$(\theta> \allowbreak 60^{\circ})$ \cite{PierreAuger:2014ucz,PierreAuger:2021qsd}, 
also named horizontal air-showers (HAS), since the electromagnetic component is largely attenuated 
in the atmosphere and thus only muons and their EM halo reach the ground.
The reconstructed total number of muons at the ground is calculated relative to the expected value 
by a reference  model  at $10^{19}~\mathrm{eV}$, giving the observable  $R_{\mu}$.
This observable is obtained by introducing a normalization factor in the reference model to fit the muon density at 
the ground to the signals measured in the surface array. 
The measured averages of $R_{\mu}$, presented in Fig. \ref{Fig:Rmu2}-left, are marginally 
compatible with predictions for iron-induced showers within the systematic uncertainty. 
However, the estimated mass composition  by means of $X_{\rm{max}}$ measurements is not 
dominated by iron, as shown in Fig.  \ref{Fig:Rmu2}-right for showers at $10^{19}~\mathrm{eV}$.
The measured number of muons is clearly larger than model predictions.

The first measurements of the muon content by AMIGA with the engineering array of UMDs 
were recently published \cite{PierreAuger:2020gxz}, and are also plotted in Fig. \ref{Fig:Rmu2}-left. 
The muon content  is compatible with iron primaries as for highly inclined showers. 
Including mass composition interpretation based on $X_{\rm{max}}$ measurements 
gives a result similar to Fig. \ref{Fig:Rmu2}-right, i.e., more muons are found in data than in simulations.
Hence, AMIGA shows a muon deficit even at lower energies, down to $3 \times 10^{17}~\mathrm{eV}$.

\section{Fluctuations of the muon number}
\label{sec:SigmaNmu}
\begin{figure}[h!]
\centering
\includegraphics[width=1.\textwidth]{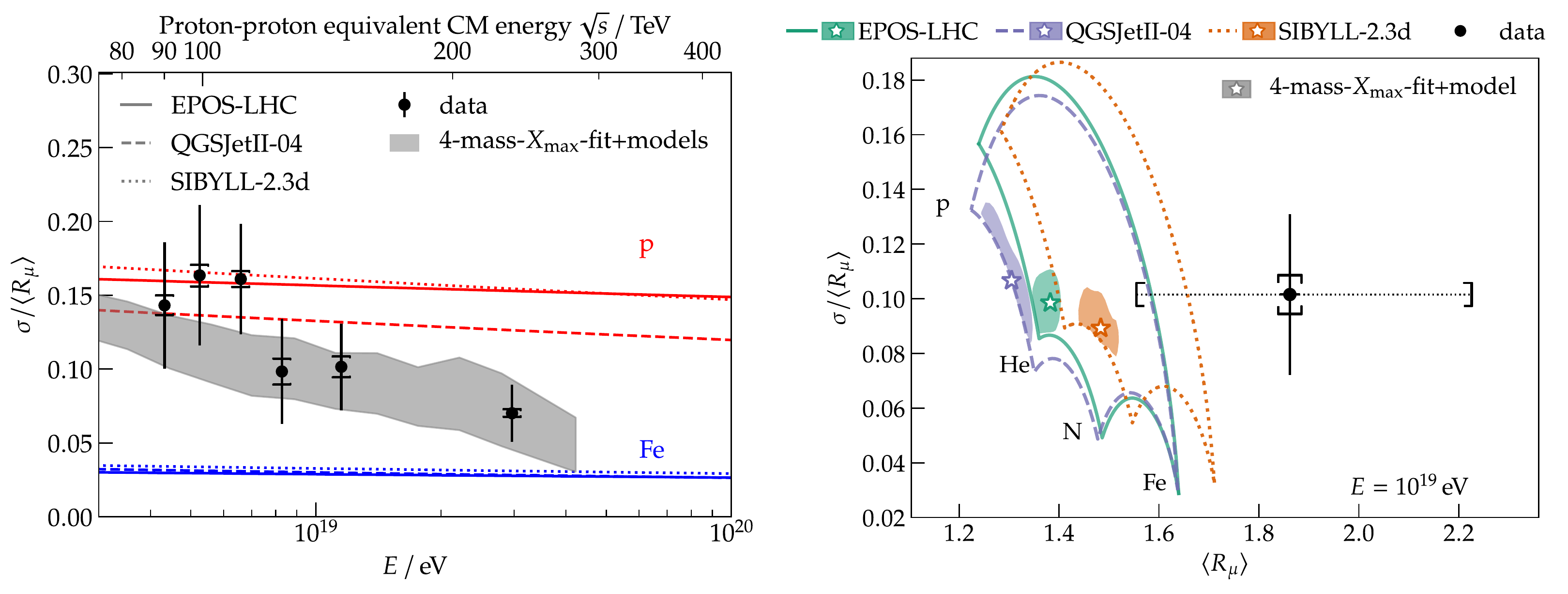}
\vspace{-7mm}
\caption{ Left: The energy evolution of the measured relative fluctuations in the number of muons
compared to simulations \cite{PierreAuger:2021qsd}.
The expected values from mass composition measurements is shown as a gray band.
Right:  Measured and simulated fluctuations and average  muon number for showers at  $10^{19}~\mathrm{eV}$ \cite{PierreAuger:2021qsd}. 
More details in the text.}
\label{Fig:SigmaNmu}
\end{figure}

A recently published result by the Pierre Auger Collaboration is the analysis of muon fluctuations \cite{PierreAuger:2021qsd}.
The measured relative fluctuations in the number of muons as a function of energy 
and the predictions from  models are shown in Fig. \ref{Fig:SigmaNmu}-left. 
The models give consistent results since the measurement falls within the expected range for  proton and  iron primaries.
Fig. \ref{Fig:SigmaNmu}-right shows the measured fluctuations and average number of
muons  at a fixed primary energy of $10^{19}~\mathrm{eV}$.  
The expected values predicted by the models for the mixture  of p, He, N, Fe 
from $X_{\rm{max}}$ measurements are indicated by the star symbols.
The shaded areas are the regions allowed by the uncertainties, while the colored 
contours are any possible mixture of those four primaries. 
So the muon fluctuations  are well reproduced by models but not the total number of muons. 
These results  point to a small muon deficit at every stage of the shower rather than a discrepancy in the first interactions.

\section{Conclusions}
\label{sec:Conclusions}
The Pierre Auger Observatory uses a hybrid system to measure the properties of
extensive air showers. Recent results have constrained the high-energy interaction models.
The direct measurements of the p-air cross section at $\sqrt{s} = 39$ TeV and 56 TeV constrain the
extrapolations of the hadronic models towards the highest energies.
On the other hand, the mass composition interpretation based on the measurements of 
the depth of shower maximum in addition to the measured muon content at ground level 
indicate a deficit in the number of muons predicted by models even at  energies down to $3 \times 10^{17}~\mathrm{eV}$. 
This result implies that the muon production mechanism is not well described by the current hadronic interaction models. 
Several models were developed to explain this deficit.
Some of them implement modifications of hadronization parameters changing the ratio of electromagnetic and hadronic components, 
while other models change the inelastic cross section or the property of the decays.
However, the analysis of muon fluctuations  points to a small muon deficit at each hadronic interaction. 
This indication disfavours exotic models describing the highest energy interactions.
The AugerPrime upgrade of the Pierre Auger Observatory will improve the sensitivity to hadronic interactions 
and mass composition, allowing to explore further hadronic interactions beyond 
the energy reached by colliders and at the forward region.

\bibliographystyle{SciPost_bibstyle} 
\bibliography{biblography.bib}

\nolinenumbers

\end{document}